\documentclass[fleqn,usenatbib]{mnras}

\usepackage[T1]{fontenc}

\DeclareRobustCommand{\VAN}[3]{#2}
\let\VANthebibliography\thebibliography
\def\thebibliography{\DeclareRobustCommand{\VAN}[3]{##3}\VANthebibliography}

\usepackage{graphicx}	
\usepackage{amsmath}	
\usepackage{amssymb}	
\usepackage{caption}    
\usepackage{subcaption} 
\usepackage{gensymb}    

\title[Gas content of field dwarf galaxies]{The present-day gas content of simulated field dwarf galaxies}

\author[G. Herzog et al.]{
Georg Herzog,$^{1}$\thanks{E-mail: g.herzog@campus.unimib.it}
Alejandro Benitez-Llambay,$^{1}$ and
Michele Fumagalli$^{1,2}$
\\
$^{1}$Dipartimento di Fisica 'G. Occhialini', Universit\`a degli Studi di Milano-Bicocca, Piazza della Scienza 3, I-20126 Milano, Italy\\
$^{2}$INAF - Osservatorio Astronomico di Trieste, via G. B. Tiepolo 11, 34143 Trieste, Italy \\
}

\date{Accepted XXX. Received YYY; in original form ZZZ}

\pubyear{2021}

\defcitealias{BenitezLlambayFrenk2020}{BL20}
\defcitealias{Benitez-Llambay2013}{BL13}

\begin{document}
\label{firstpage}
\pagerange{\pageref{firstpage}--\pageref{lastpage}}
\maketitle

\begin{abstract}
We examine the gas content of field dwarf galaxies in a high-resolution cosmological simulation. In agreement with previous work, we find that galaxies inhabiting dark matter haloes with mass below a critical value, $M_{200} \lesssim M_{\rm crit} \approx 5\times 10^{9} \ M_{\odot}$, are quiescent at the present day. The gas content of these galaxies is thus insensitive to feedback from evolving stars. Almost half of these quiescent systems today have gas masses much smaller than that expected for their mass. We find that gas-deficient galaxies originate from 1) past interactions with massive hosts, in which a dwarf loses gas and dark matter via tidal and ram-pressure forces; and 2) from hydrodynamic interactions with the gaseous filaments and sheets of the cosmic web, in which a dwarf loses gas via ram-pressure. We refer to these systems as ``flybys'' and ``COSWEBs''. Flybys locate in high-density regions, tracing the location of the most massive galaxies in the simulation. In contrast, COSWEBs are dispersed throughout the volume and trace the cosmic web. For sub-critical systems, $M_{200} < M_{\rm crit}$, the fraction of COSWEB galaxies can be as high as $35 \%$, and much higher for flybys, which make up 100 per cent of the galaxies with $M_{200}<3\times 10^8 \ \rm M_{\odot}$. The deficit of gas caused by these mechanisms may preclude the detection of a large fraction of field dwarfs in future \ion{H}{I} surveys. For galaxies inhabiting halos with mass $M_{200} > M_{\rm crit}$, we find that cosmic web stripping, on average, shuts down star formation in more than $70\%$ of the affected systems.
\end{abstract}




\section{Introduction}
\label{Sec:Introduction}

Within the $\Lambda$ Cold Dark Matter ($\Lambda$CDM) paradigm of structure formation, galaxies form from gas that condenses to turn into stars in the centre of gravitationally-bound dark matter haloes~\citep{White1978}. The amount of baryons that reside in these haloes depends, however, on the considered mass and cosmic time scales.

At early times, for example, before the epoch of cosmic reionization, dark matter haloes sufficiently massive to overcome the entropy of the intergalactic medium can accumulate gas copiously, reaching gas mass fractions similar to the cosmic mean, $f_{\rm b} \sim \Omega_{\rm b} / \Omega_{\rm m}$~\citep[e.g.][]{Naoz2009}. The gas in these haloes is, however, quickly pressurised through shocks and cannot collapse further to the centre of the majority of systems to form stars. This is because galaxy formation at early times proceeds predominantly in atomic cooling (AC) haloes, i.e., those haloes with virial\footnote{We identify virial quantities with a $200$ subscript. These quantities correspond to those measured within a sphere for which the mean enclosed density is 200 times the critical density of the Universe, $\rho_{\rm crit} = 3H^2/8\pi G$.} temperature $T_{200} \gtrsim 10^{4} \ K$,\footnote{The virial temperature is defined as $T_{200} = 36 (V_{200}/\rm km \ s^{-1})^2 \ K$, where $V_{200}$ is the halo virial circular velocity, $V_{200} = (G M_{200} / R_{200} )^{1/2}$, and we assume a mean molecular weight $\mu = 0.6$.} for which gas can lose its pressure support via radiative cooling~\citep[see, e.g., a review by][]{Bromm2011}. Once galaxies form in the centre of these haloes, the efficient expulsion of gas by baryonic processes associated with stellar evolution, including radiative, thermal, and kinetic feedback, affects their baryon content dramatically, and numerical hydrodynamical simulations are required to follow their gas content in detail, which is largely determined by the net balance between gas infall and outflows.

At later times, after the epoch of reionization ($z<z_{\rm re}$), the presence of the ionizing UV background (UVB) radiation field prevents gas accretion onto haloes less massive than a present-day mass, $M_{200} \lesssim M_{\rm crit} \approx 5 \times 10^{9} \ M_{\odot}$, a mass-scale that exceeds that imposed by the AC limit by roughly a factor of 5.\footnote{Note that at $z=0$, the AC limit corresponds to halo virial mass $M_{200} \sim 10^9 \ M_{\odot}$.} By cutting out their gas supplies, the photoheating associated with the UVB inhibits further star formation in the already existing galaxies that inhabit these low-mass haloes~\cite[e.g.][and references therein]{Efstathiou1992, Thoul1996, Quinn1996, Weinberg1997, Babul1992, Bullock2000, Benson2002}, and prevents the formation of galaxies in starless haloes less massive than $M_{\rm crit}$. Although the majority of baryons are indeed pushed out from these low-mass systems, an increasingly small fraction of gas remains bound to dark matter haloes down to a present-day virial mass, $M_{200} \gtrsim 10^6 \ M_{\odot}$, below which dark matter haloes become baryon-free~\citep[][hereafter BL20]{BenitezLlambayFrenk2020}. The gas content of haloes in the mass range $10^6 \lesssim M_{200} / M_{\odot} \lesssim 5\times 10^9$ is thus established by the balance between the gravitational potential and the pressure of the intergalactic medium~\citep{Rees1986, Ikeuchi1986, BenitezLlambay2017}, and it is insensitive to the stellar content of the halo because the gravitational contribution of stars in these low-mass systems can be largely neglected. On the other hand, the baryon content of haloes with virial mass $M_{200} > M_{\rm crit}$, for which gas cannot remain in hydrostatic equilibrium, is affected by feedback associated with star formation, and simulations are required to track the gas content of these haloes.

It is thus clear that the baryon content of galaxy haloes is largely regulated by the balance between gas infall and outflows, but only for haloes more massive than the AC limit before cosmic reionization, and for haloes more massive than $M_{\rm crit}$ afterwards. In the present-day halo mass range, $10^6 \lesssim M_{200} / M_{\odot} \lesssim M_{\rm crit} \approx 5\times 10^9$, for which gas is stable against gravitational collapse (and therefore unable to foster star formation), the small amount of gas left inside these haloes remains in hydrostatic equilibrium~\citepalias{BenitezLlambayFrenk2020}. 

The previous picture strictly applies to systems that form and evolve in relative isolation, as the gas content of galaxy haloes may be affected, in turn, by external environmental processes. For example, it is well known that hydrodynamic and gravitational interactions can remove gas from galaxy haloes and cut off their gas supplies, and similarly to the previous mechanisms, their impact on the gas content is exacerbated for low-mass haloes. These interactions affect predominantly satellite galaxies that fall into groups and clusters~\citep[e.g.,][and references therein]{Gunn1972, Moore1996, Abadi1999, McCarthy2008}, and are thought to be responsible for the environmental dichotomy observed within our Local Group~\citep[e.g.,][]{VanDenBergh1994}, in which the majority of satellites of the Milky Way (MW) and Andromeda (M31) are devoid of gas and are not forming stars at the present day, whereas the fraction of gas-rich dwarfs increases dramatically further out from the host galaxies~\citep[see, e.g., a recent compilation by][]{Putman2021}. 

The existence of the environmental dichotomy in gas mass observed today in the Local Group constitutes perhaps the strongest argument in favour of the idea that isolated galaxies represent true examples of how galaxies form and evolve in the absence of environmental processes. However, the existence of nearby dwarfs in relative isolation that exhibit particularly low gas-to-stellar mass ratios~\cite[e.g.][]{Karachentsev2014} seem to challenge this idea. Moreover, using high-resolution numerical simulations, \cite[][hereafter BL13]{Benitez-Llambay2013} have shown that isolated dwarf galaxies may be subject to ``cosmic-web stripping'', a process that can either suppress star formation by removing gas from dwarf galaxy haloes in the case of strong interactions between the galaxy gaseous halo and the cosmic web or reignite star formation by compressing the gas for the case of weaker interactions~\citep[e.g.][]{Wright2019, Genina2019}. The efficiency of this process in shaping the present-day gas content of the cosmological population of dwarfs is, however, difficult to assess, as the simulations considered by these authors encompass, in all cases, very limited (and biased) cosmological volumes.

The goal of our paper is thus to revisit the importance of the environment, and in particular of cosmic-web stripping, in establishing the present-day gas content of isolated dwarf galaxies. To this end, we use a hydrodynamic simulation with sufficiently high resolution to track the gas content of dwarf galaxy haloes and with sufficient large volume to understand the importance of this process over cosmological scales. We describe our simulation and galaxy sample in Sec.~\ref{Sec:Methods}. We then present and discuss our results in Sec.~\ref{Sec:Results} and Sec.~\ref{Sec:Discussion}, respectively, and conclude with a summary of our results in Sec.~\ref{Sec:Conclusions}.

\section{Methods}
\label{Sec:Methods}

\subsection{The simulation}
We use a Smoothed-Particle-Hydrodynamics (SPH) cosmological simulation evolved with the {\tt P-Gadget3} code~\citep{Springel2005} and the {\tt EAGLE} model of galaxy formation~\citep{Schaye2015, Crain2016}. The simulation is the same introduced by \citetalias{BenitezLlambayFrenk2020} and consists of a random realisation of a periodic cubic volume of side length $20 \rm \ Mpc$ filled with $1024^3$ gas and dark matter particles. The gas and dark matter particle mass are, respectively, $m_{\rm gas}\approx 4.5\times 10^4$ M$_{\odot}$, and $m_{\rm dm}\approx 2.4\times 10^5$ M$_{\odot}$, which ensures the simulated volume is at the mean density of the Universe. The initial conditions were carried out at redshift, $z=127$, with the publicly-available code {\tt MUSIC}~\citep{Hahn2011}. The Plummer-equivalent gravitational softening, $\epsilon$, adopted in our simulation never exceeds $1\%$ of the mean interparticle separation. This gives $\epsilon \sim 195 \rm \ pc$ for both the gas and the dark matter particles. This value is much smaller than the radius below which the 2-body relaxation timescale equals the age of the Universe for our simulation~\citep{Power2003}. The structure of the haloes in our simulation is thus not limited by the choice of $\epsilon$, but by the number of particles. We list here the simulation details relevant to our study, and we refer the reader to the original papers for further details.

Star formation is implemented in the simulation by turning gas particles into stars at the same Kennicutt–Schmidt rates adopted in the EAGLE simulations, but only for gas particles that exceed a density threshold $\rho_{\rm th} = 1.0 \rm \ cm^{-3}$. This high threshold ensures that the gas is self-gravitating in the centre of the haloes before turning into stars~\citep{Benitez-Llambay2019}. Unlike the original {\tt EAGLE} model, the density threshold for star formation in our simulation does not depend on metallicity. The simulation includes gas cooling and heating by the external UVB, which is turned on at the redshift of reionization, $z_{\rm re} = 11.5$, and corresponds to that of \cite{Haardt2001}. We assume cosmological parameters consistent with early Planck results~\citep{Planck2014}. 

Our mass resolution is $\sim 100$ higher than that of the original EAGLE suite, making our simulation particularly well-suited to resolve dwarf galaxy haloes with several hundred DM particles. The star particles have a mass identical to that of the gas particles, imposing a minimum galaxy mass in our model, $M_{\rm \star} \sim 4.5 \times 10^4 \ M_{\odot}$. Dark matter haloes are identified in the simulation with the {\tt HBT+} code~\citep{Han2018}, which uses a catalogue of Friend-of-Friends (FoF) haloes constructed with a percolation length $b=0.2$, in units of the mean interparticle separation, to carry out an unbinding procedure based on the gravitational binding energy of the particles. {\tt HBT+} returns a catalogue of gravitationally-bound haloes, together with a list of particles (dark matter, stars, and gas) bound to each halo. {\tt HBT+} also classifies haloes as either ``centrals'' or ``satellites'' based on whether the halo is the more massive substructure of the FoF group or not, respectively.  We refer the reader to the original {\tt HBT+} paper for extensive details on the algorithm.

\begin{figure}
    \centering
    \includegraphics[width=\columnwidth]{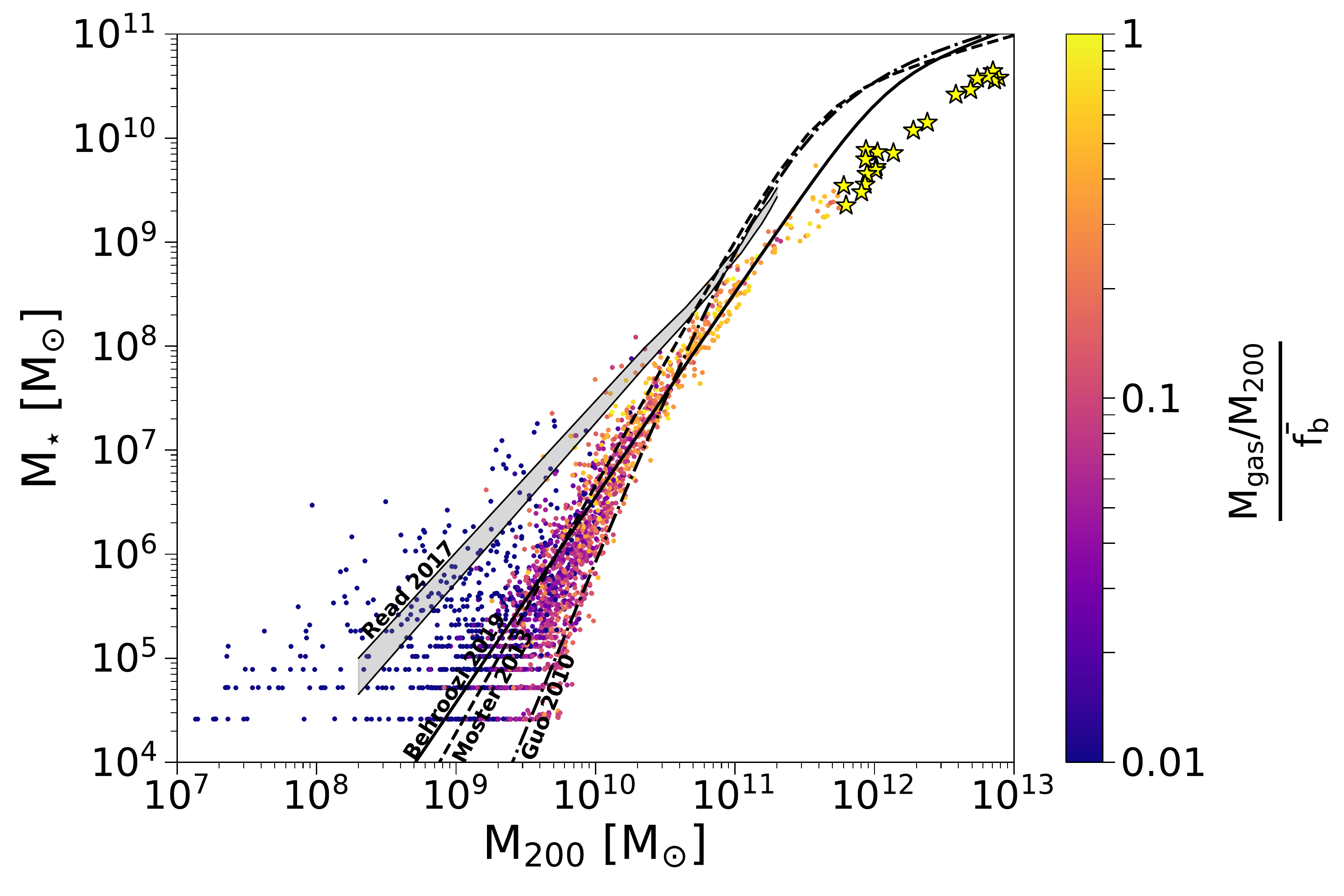}
    \caption{Present-day stellar mass, $M_{\rm \star}$, as a function of virial mass, $M_{200}$, for the isolated galaxies identified in the simulation. Galaxies are coloured according to their virial gas mass fraction, $M_{\rm gas}/M_{200}$, relative to the universal baryon fraction, $\bar f_{\rm b} = \Omega_{b}/\Omega_{m}$. The various lines display different abundance matching expectations, as labelled. Yellow stars show the galaxies inhabiting the 20 most massive haloes of the simulation.}
    \label{Fig:Stellar-vs-HaloMass}
\end{figure}

\begin{figure*}
    \centering
    \includegraphics[width=\textwidth]{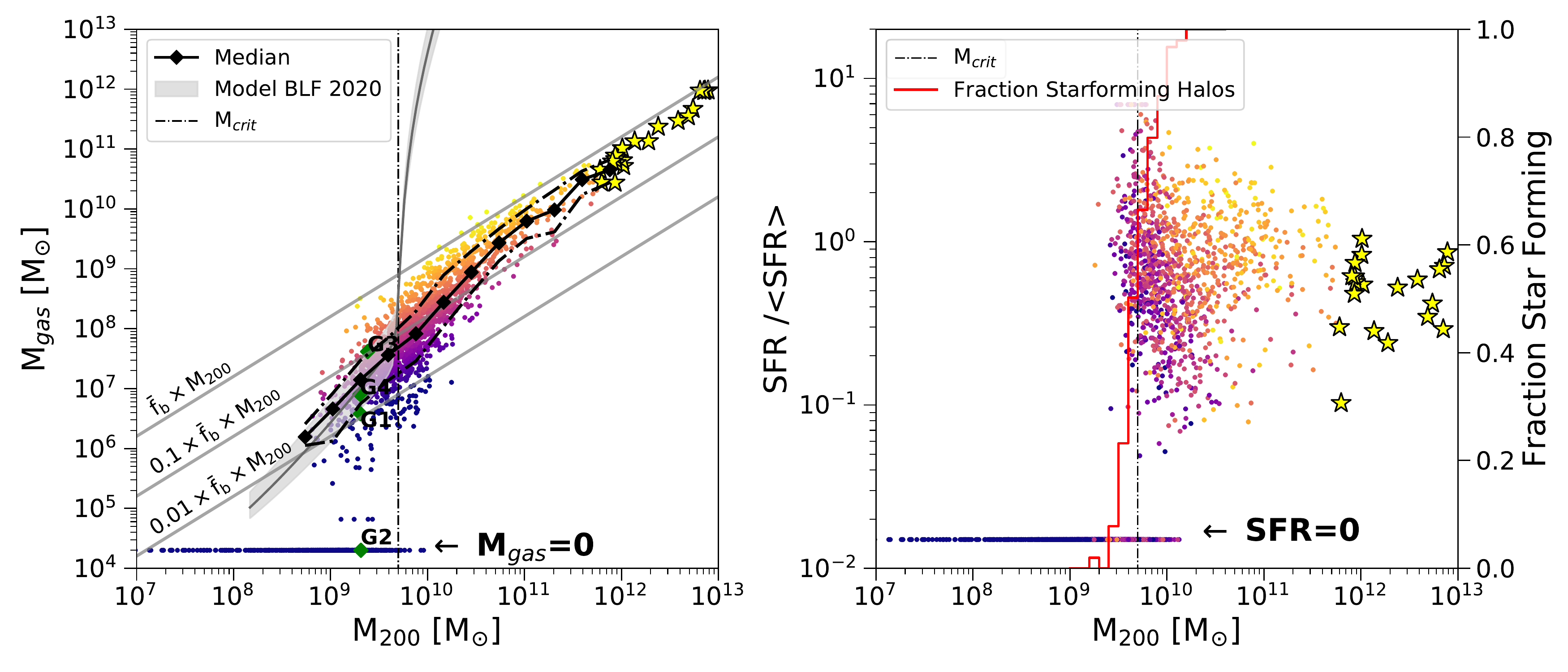}
    \caption{The left panel shows the present-day gas mass as a function of halo mass for our galaxy sample. We display galaxies with $M_{\rm gas}=0$ with an arbitrarily low value (horizontal arrow). Galaxies are coloured as in Fig.~\ref{Fig:Stellar-vs-HaloMass}, i.e., according to their gas mass fraction relative to the universal baryon fraction. The solid black line indicates the running median of the distribution for galaxies with $M_{\rm gas}>0$, and the dot-dashed lines indicate the 16-84th percentiles, as measured in bins equally spaced logarithmically in mass (black symbols). The grey line and shaded region show the gas mass that results from applying the \citetalias{BenitezLlambayFrenk2020} model. The vertical dot-dashed line indicates the critical virial mass above which gas cannot remain in hydrostatic equilibrium according to this model. The oblique lines display the universal baryon fraction, $\bar f_{\rm b}=\Omega_{b}/\Omega_{m}$, 10 per cent, and 1 per cent of this value. The green diamonds indicate four example galaxies (G1, G2, G3, and G4) that inhabit halos of the same mass today but contain very different gas masses. The right panel shows the star formation rate (SFR), in units of the past average, as a function of present-day halo mass. The red histogram shows the fraction of systems for which SFR$ > 0$ (scale on the right). We display galaxies with $\rm SFR=0$ with an arbitrarily low value (horizontal arrow). Galaxies less massive than the~\citetalias{BenitezLlambayFrenk2020} critical mass are quiescent, as expected. In both panels, the yellow stars indicate the 20 most massive systems in the simulation.}
    \label{Fig:Fig2_MgasMhaloAndSFRMhalo}
\end{figure*}

\subsection{Sample selection}

Our goal is to examine the current gas content of isolated dwarf galaxies. Therefore, we restrict our galaxy sample to systems labelled as ``centrals'' by {\tt HBT+} at the present day. As opposed to satellite systems, central galaxies do not reside within the virial radius of more massive counterparts, by definition. In addition to the isolation criterion, we require the haloes to contain a galaxy, i.e., to have at least one bound stellar particle. These two conditions yield a population of central galaxies resolved with more than $\sim 100$ dark matter particles. After a detailed inspection of the galaxies' individual mass accretion histories, we removed 14 galaxies that were wrongly identified by {\tt HBT+}. Our selection criteria thus result in 2268 galaxies that span the stellar mass range, $10^4 \lesssim M_{\rm \star} / M_{\odot} \lesssim 10^{11} \ M_{\odot}$, and the halo mass range, $10^{7} \lesssim M_{200}/M_{\rm \odot} \lesssim 10^{13}$. As shown in Fig.~\ref{Fig:Stellar-vs-HaloMass}, the stellar vs halo mass relation of the simulated galaxies is broadly consistent with abundance matching (AM) expectations from~\cite{Moster2013},~\cite{Read2017},~\cite{Behroozi2019} and~\cite{Guo2010} but only at low masses. Our simulated galaxies form too few stars compared to AM at high masses. This mismatch, however, does not preclude the results that follow, as we are interested in dwarf galaxies only.

\begin{figure*}
    \centering
    \includegraphics[width=\textwidth]{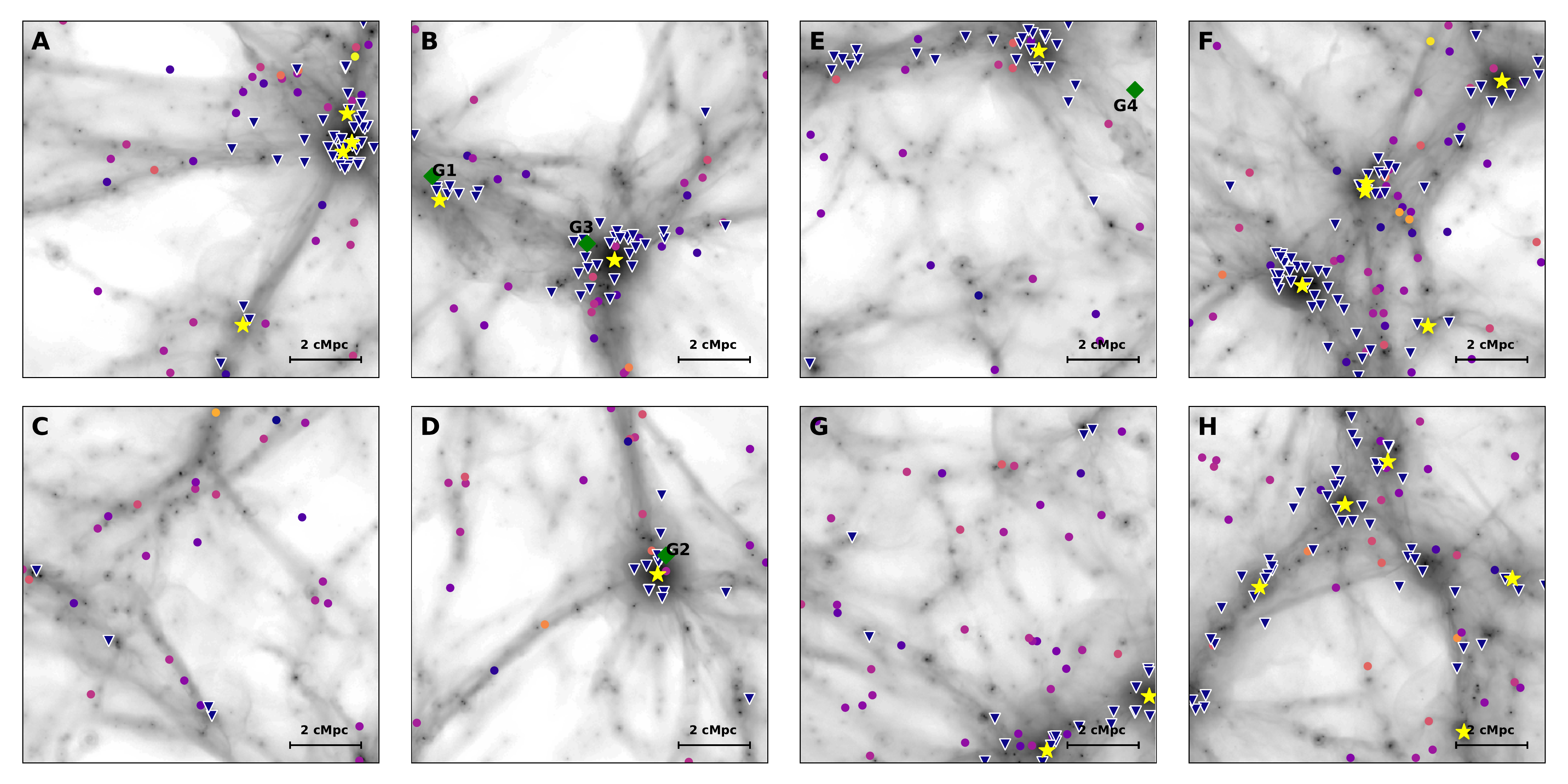}
    \caption{Spatial distribution of simulated galaxies spanning the narrow range in virial mass, $10^9 < M_{200} / M_{\odot} <3 \times 10^9$, in cubic regions of 10 Mpc side length equally spaced to cover the entire simulated volume. Panels A, B, C, and D may be combined to form a slice spanning the coordinates range, $(x,y,z)=(0-20,0-20,0-10)$ Mpc, whereas panels E, F, G, and H may be combined to cover the remaining half of the box, i.e., $(x,y,z)=(0-20,0-20,10-20)$ Mpc. As in Fig. \ref{Fig:Fig2_MgasMhaloAndSFRMhalo}, we colour galaxies according to their present-day gas fraction relative to the universal baryon fraction. Yellow stars indicate the location of the 20 most massive galaxies in the simulation. The colour map in the background displays the projected gas density. Gas-deficient galaxies (blue triangles with white edges) cluster either along the filaments of the cosmic web or close to massive systems, in sharp contrast with the other galaxies in the same halo mass range. The strongly clustered distribution of gas-deficient galaxies thus points to the role of the environment in shaping their present-day gas content.}
    \label{Fig:Fig3_OktantsHaloMassBin}
\end{figure*}

\section{Results}
\label{Sec:Results}

\subsection{Present-day gas content of isolated simulated galaxies}

The left panel of Fig.~\ref{Fig:Fig2_MgasMhaloAndSFRMhalo} shows the present-day bound gas mass of our galaxy sample as a function of virial mass. Galaxies are coloured according to their gas mass fraction, $f_{\rm b} = M_{\rm gas}/M_{200}$, relative to the universal mean, $\bar f_{\rm b} = \Omega_{b} / \Omega_{m}$ (shown by the top solid oblique line in the same panel). The black symbols indicate the running median and the 16-84th percentiles of the distribution (but only for galaxies for which $M_{\rm gas} > 0$). Except for a few systems, most galaxies in our simulation do not retain the universal baryon budget within their virial boundaries. Indeed, more than half of the systems at all masses have lost more than $2/3$ of their baryons at $z=0$\footnote{Note that the contribution of the stars can be largely neglected to the virial baryonic budget (see Fig.~\ref{Fig:Stellar-vs-HaloMass}).}. For massive systems ($M_{200} >> 10^{10} \ M_{\odot}$), the loss of baryons is largely due to the effect of the efficient supernova feedback~\citep[see, e.g.,][and references therein for a recent study on the baryon content of massive haloes in the EAGLE simulations]{Wright2020,Mitchell2022}. For less massive systems, the loss of baryons is exacerbated by their shallower potential wells and by the presence of the external UVB radiation field~\citep[e.g.][]{Okamoto2008}. 

Supernova feedback is responsible, too, for the removal of baryons in low-mass systems, but only for haloes more massive than $M_{\rm crit}\approx 5 \times 10^{9} \ M_{\odot}$, for which gas becomes self-gravitating in the centre and forms stars~\citep[see Sec.~\ref{Sec:Introduction}, and the work of][]{Pereira-Wilson2022}. For haloes with mass $M_{200} \lesssim M_{\rm crit}$ this is not the case, as we demonstrate in the right-hand panel of the same figure, where we show the star formation rate (SFR) of our galaxy sample, in units of the past average, as a function of present-day halo mass. The star formation rate is simply defined as $\Delta M / \Delta t$, i.e., the ratio between the mass of stars formed within the galaxy radius, $R_{\rm gal} = 0.2 \times R_{200}$, in the last $\Delta t=2 \rm \ Gyrs$, and $\Delta t$. The past average is defined as $<\rm SFR>=M_{\rm gal} / t_{\rm H}$, where $t_{\rm H}$ is the Hubble time. We show galaxies that have not formed stars in this time interval with an arbitrarily low value indicated by the arrow at the bottom right of the same panel. The red solid line shows the fraction of galaxies, as a function of halo mass, for which ${\rm SFR}>0$ (scale on the right). This rough definition of SFR indicates that, on average, massive systems in the simulation have formed stars at a constant rate in the past few billion years. Galaxies inhabiting haloes with masses $M_{200} \lesssim M_{\rm crit}$, on the other hand, have not formed stars in the same time interval. Thus, the systematic reduction in their gas content compared to more massive systems cannot be due to the effect of supernova feedback. The transition between star-forming and quiescent dwarfs occurs close the~\citetalias{BenitezLlambayFrenk2020} critical mass (shown by the vertical dot-dashed line), below which gas remains in hydrostatic equilibrium.

The thin grey solid line in the left-hand side panel of Fig.~\ref{Fig:Fig2_MgasMhaloAndSFRMhalo} shows the gas mass as a function of halo mass that results from the~\citetalias{BenitezLlambayFrenk2020} model. This model derives the gas mass of a  dark matter halo after assuming that the gas is in thermal equilibrium with the external UVB and in hydrostatic equilibrium with a Navarro-Frenk-White (NFW; \citealt{Navarro1996, Navarro1997}) halo. To derive the gas mass, the model further assumes that the gas pressure sufficiently far from the system is that of the intergalactic medium at the mean density of the Universe, $\bar \rho_{\rm b}$. The grey shaded region shows the expected gas mass for haloes embedded in an environment three times over/under-dense (top/bottom envelopes). The agreement between the median gas mass of galaxies with $M_{\rm gas}>0$ and the model demonstrates that the gas mass of galaxies inhabiting haloes with mass, $M_{200} \lesssim M_{\rm crit}$, is indeed established by the effect of the external photoheating background,\footnote{We note that the gas mass of subcritical halos depends weakly on the assumed photoheating background~\citep[see, e.g.,][]{BenitezLlambay2017}.} and offers the physical explanation as to why these galaxies have not been forming stars for such a long time. Interestingly, some of these quiescent galaxies do not contain gas bound to them (see the left panel in Fig.~\ref{Fig:Fig2_MgasMhaloAndSFRMhalo}), or when they do, their gas content departs significantly from that expected for isolated galaxies. The gas mass of these galaxies is thus neither established by the external UVB nor by supernova-driven winds. 

Defining gas-deficient systems as galaxies whose present-day gas fraction is $M_{\rm gas}/M_{200} < 0.01 \times \bar f_{\rm b} = 0.01 \times \Omega_{b}/\Omega_{m}$, does a good job at splitting our galaxy sample between normal systems (those galaxies whose gas mass is well-understood in terms of the~\citetalias{BenitezLlambayFrenk2020} model), and gas-deficient systems. Therefore, we shall adopt this definition in what follows. We address the origin of these gas-deficient galaxies next.

\begin{figure*}
    \includegraphics[width=\textwidth]{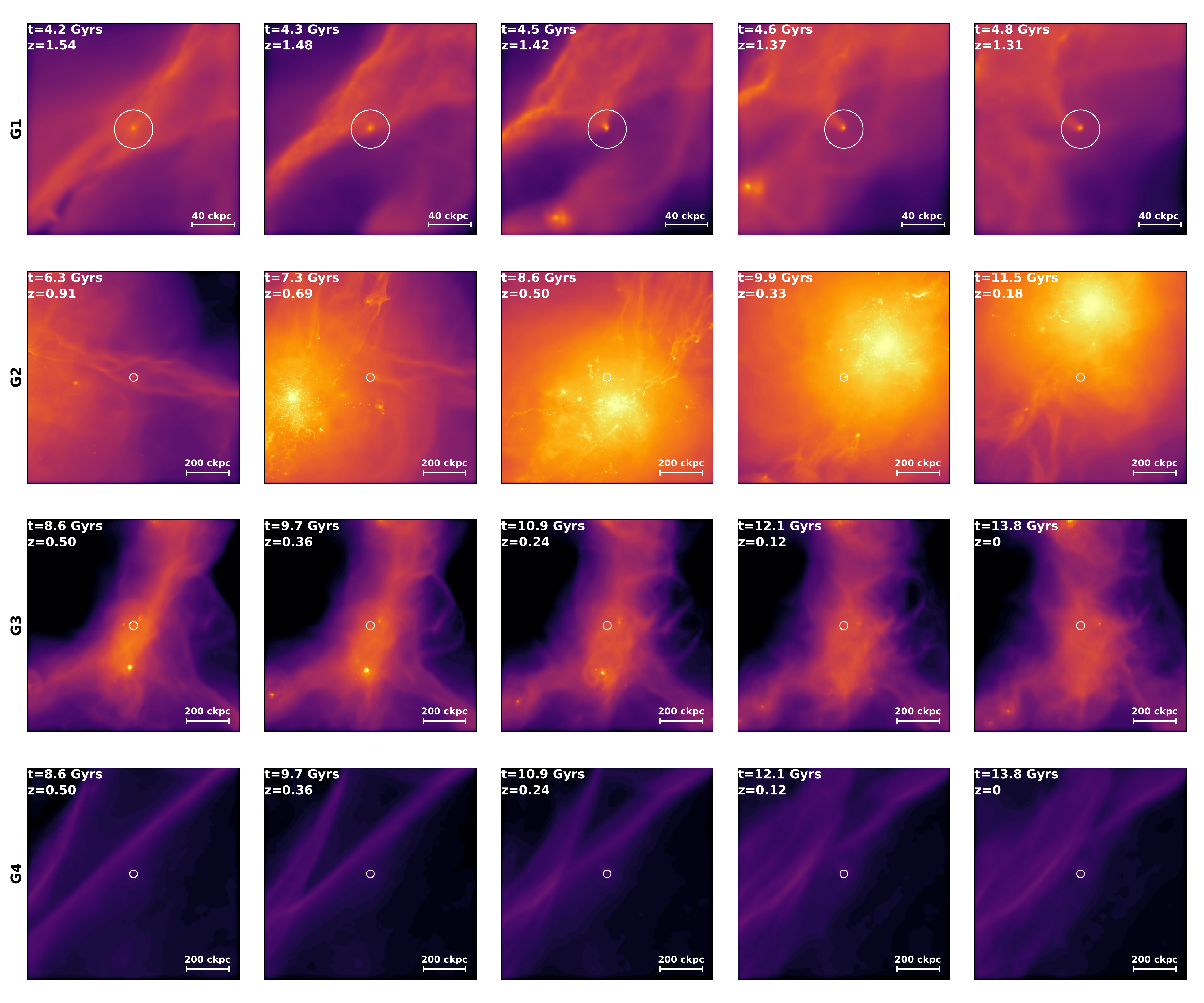} 
    \caption{Time evolution of 4 individual simulated dwarf galaxies. Different rows show the projected gas density around each dwarf at various times. The white circle indicates the virial radius of the galaxy. The first row shows the evolution of G1, which loses its gas by cosmic web stripping. The second row displays G2, which, although an isolated system today, approached a massive system in the past and lost gas and dark matter via tidal and ram pressure forces. The third row shows the evolution of G3, a galaxy embedded in a dense but otherwise isolated environment, in contrast to G4 (fourth row), which evolves in a less dense but isolated environment.}
    \label{Fig:Fig4_4Examples}
\end{figure*}

\subsection{Role of environment in establishing the present-day gas content of field dwarf galaxies}

Fig.~\ref{Fig:Fig3_OktantsHaloMassBin} provides clues to the origin of the population of isolated, quiescent, and gas-deficient galaxies. Here we show the spatial distribution of dwarfs inhabiting the narrow range in halo mass, $10^{9} \lesssim M_{200} / M_{\odot} \lesssim 3\times 10^{9}$, in cubic regions of 10 Mpc side length taken from the parent volume. We colour galaxies as in Fig.~\ref{Fig:Fig2_MgasMhaloAndSFRMhalo} (i.e., according to their gas fraction relative to the cosmic mean). The colour map in the background shows the projected gas density.

Interestingly, gas-free dwarfs (blue triangles with white edges) cluster preferentially in the surroundings of massive systems (shown by yellow stars), albeit further away from their virial radius, by construction. In contrast, the rest of the galaxies reside further away from the overdense regions depicted by the most massive systems in the volume. Consider, for example, panel C of Fig.~\ref{Fig:Fig3_OktantsHaloMassBin}, where we can hardly identify gas-free haloes. This particular region of our simulation lacks massive galaxies and prominent large-scale gaseous structures. A similar analysis applies to panel G, where only very few gas-free dwarfs are found at the bottom right corner of the volume, a region that contains two massive galaxies and a prominent filament that connects them. If we focus on panel H, we see that gas-free galaxies cluster around the five massive galaxies present in this volume or depict the location of dense gaseous filaments. The rest of the normal galaxies in terms of their gas fraction, i.e., those whose gas mass is roughly consistent with simple analytic expectations, lie further away from overdense regions. A similar analysis holds to the remaining volumes. Thus, it is tempting to ascribe the dichotomy in the gas mass to the characteristic environment surrounding each dwarf. In the next section, we analyse four individual examples in detail.

\subsection{Individual examples}

To demonstrate how the different environments, together with the past evolution of field galaxies, shape their current gas mass, we shall consider four individual examples that exhibit disparate present-day gas contents. These are labelled G1, G2, G3, and G4. As shown by the green diamonds in the left panel of Fig.~\ref{Fig:Fig2_MgasMhaloAndSFRMhalo}, these galaxies are selected to inhabit dark matter haloes with virtually the same mass today but with very different gas mass. 

\subsubsection{Past encounter with a massive system}

First, consider galaxy G2, which inhabits a halo currently devoid of gas. This galaxy locates, similarly to the rest of the gas-free galaxies, in the surroundings of a more massive companion (see panel D in Fig.~\ref{Fig:Fig3_OktantsHaloMassBin}). The second row from the top in Fig.~\ref{Fig:Fig4_4Examples}, which shows the evolution of G2 since redshift, $z=0.91$, makes it clear that the galaxy formed in relative isolation until $z\sim 0.5$, when it encountered a massive companion. This close encounter is responsible for removing, via both ram-pressure and tidal stripping, the gas content of G2. G2 is thus an example of a system usually referred to as ``flyby'' or ``backsplash'' galaxy in the context of galaxy clusters~\cite[see, e.g.,][for earlier discussions on these galaxies]{Barlogh2000, Mamon2004, Gill2005}, but which are known to be associated with smaller systems as well~\cite[e.g.][]{Sales2007, Knebe2011}. These are galaxies that have interacted with a  massive system in the past and are now on very eccentric orbits that may reach apocentric distances several virial radii away from their host.

\begin{figure}
    \includegraphics[width=\columnwidth]{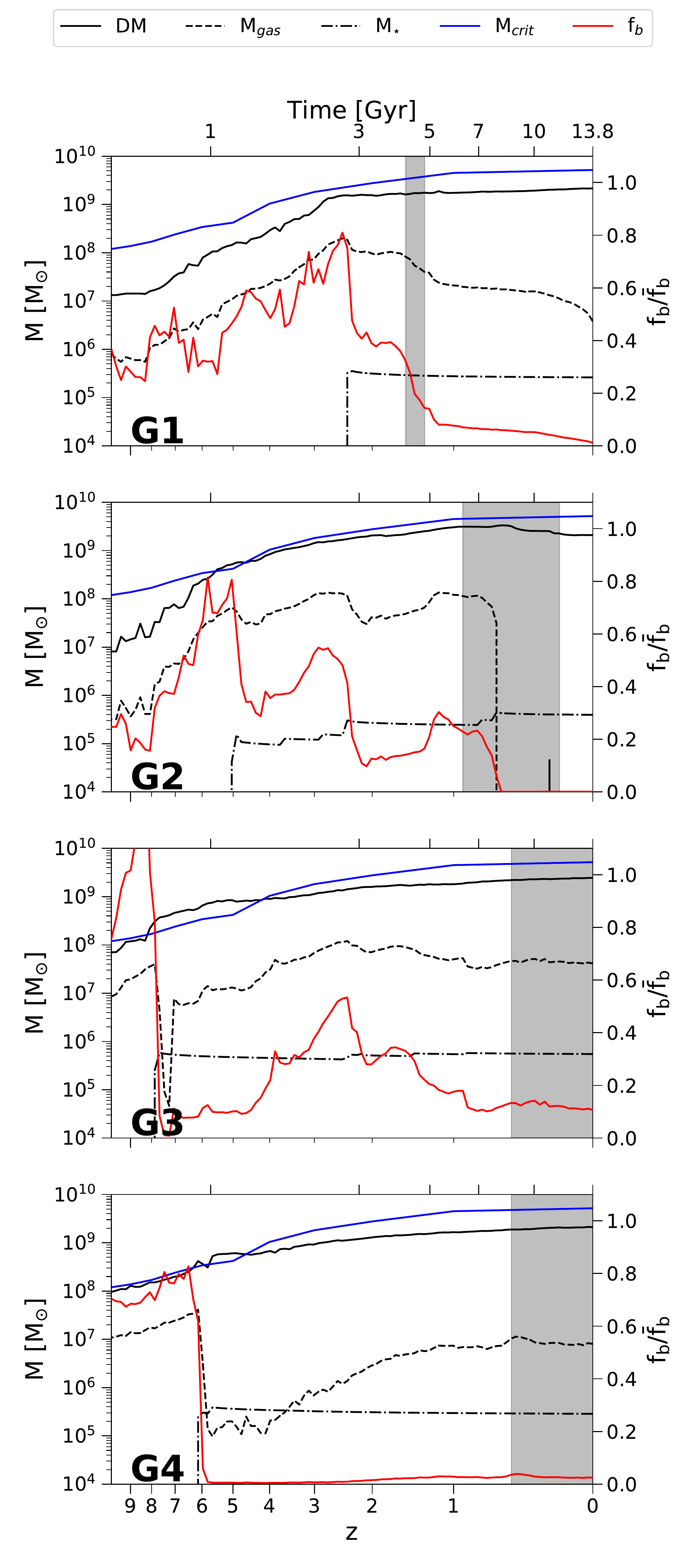} 
    \caption{Mass assembly histories of the four individual dwarfs shown in Fig.~\ref{Fig:Fig4_4Examples}. The black-solid, dashed, and dot-dashed lines show the dark matter (DM), gas, and stellar mass, respectively. The blue line corresponds to the critical mass for the onset of star formation, as predicted by the \citetalias{BenitezLlambayFrenk2020} model. The red line shows the baryon fraction, $M_{\rm gas} / M_{200}$, in units of the universal baryon fraction (scale on the right). The grey shaded area spans the time interval shown for each dwarf in Fig. \ref{Fig:Fig4_4Examples}.}
    \label{Fig:Fig9_MassAccretionHistories}
\end{figure}

We show the evolution of G2 quantitatively in the second panel of Fig.~\ref{Fig:Fig9_MassAccretionHistories}. The dark matter, gas and stellar mass are displayed by the black solid, dashed, and dot-dashed lines, respectively. The red line indicates the baryon fraction in units of the cosmic mean (scale on the right). The grey shaded region displays the time interval shown in Fig.~\ref{Fig:Fig4_4Examples}. As G2 forms stars early on, the combined effect of supernova feedback and the UVB removes roughly 80 per cent of the baryons from the system. As G2 approaches the massive companion at around $z=1$, it starts losing dark matter and the remaining gas. Some of the gas compresses during the interaction, leading to the formation of stars. G2 thus ends up being devoid of gas at $z=0$. We conclude that the interaction with a massive companion at around $z=1$ is responsible for G2's extreme deficit of gas relative to the rest of the dwarfs in the similar virial mass range. 

\subsubsection{Cosmic Web Stripping}

We now focus on galaxy G1, which inhabits a gas-deficient halo whose gas mass departs significantly from the value expected for its halo mass (see Fig.~\ref{Fig:Fig2_MgasMhaloAndSFRMhalo}). Unlike G2, galaxy G1 is much further away from a massive companion than galaxy G2. A close inspection of the evolution of this galaxy reveals the origin of its current deficit of gas compared to other galaxies of similar virial mass. The top row of Fig.~\ref{Fig:Fig4_4Examples} shows that galaxy G1 formed in relative isolation until it was swept by a gaseous filament or sheet at $z\sim 1.5$. The ram pressure exerted on G1 by this gaseous structure removes most of G1's gas, which is left in a prominent tail behind the galaxy. Note, however, that G1 does not lose all its gas. Even after the interaction, a small amount of dense gas remains deep in the centre of G1, which is however unable to form stars. G1 is thus an example of a galaxy undergoing cosmic web stripping, as described by~\citetalias{Benitez-Llambay2013}. 

In the top panel of Fig.~\ref{Fig:Fig9_MassAccretionHistories} we show the evolution of the dark matter, gas, and stellar content of G1. 

The gas mass fraction of G1 is roughly consistent with the universal mean, $\bar f_{\rm b} = \Omega_{b}/\Omega_{m}$, prior to cosmic reionization. However, the gas is quickly pushed away from the halo as the universe undergoes reionization and the interstellar medium is photoheated. This is expected, as G1 inhabits a dark matter halo less massive than the~\citetalias{BenitezLlambayFrenk2020} critical mass (shown by the blue line). As G1 approaches a dense filament, its gas compresses, enabling G1 to form stars even though G1 inhabit a halo less massive than the critical mass~\cite[see][for similar examples]{Wright2019, Pereira-Wilson2022}. This star formation episode is responsible for the sudden loss of baryons observed between $z=3$ and $z=2$, in which G1 loses almost half of its baryons. As G1 crosses the filament at around $z\sim 2$, it loses more than $3/4$ of the remaining gas. The lack of star formation after $z\sim 2$ indicates that the loss of gas after this time is not associated with supernova feedback but with the encounter of the dense filament or sheet of the cosmic web seen in Fig.~\ref{Fig:Fig4_4Examples}. After the relatively abrupt loss of gas during the interaction with the dense filament, G1 continues to lose gas steadily as it moves through the ambient gas until the present day.

\subsubsection{The impact of the ambient density}

Finally, galaxies G3 and G4 have gas masses roughly consistent with that expected for their halo mass, but G3 has retained more gas than G4 to the present day. As shown in Fig.~\ref{Fig:Fig4_4Examples}, the difference between the two galaxies may be ascribed to their environment. Both dwarfs have evolved in relative isolation, away from the massive galaxies of the simulated volume. However, G3 resides in a denser environment than that of G4\footnote{The colour scale is the same for G3 and G4. Therefore, it is possible to visually asses the relative difference in the ambient density each dwarf resides.}. 
As shown by~\citetalias{BenitezLlambayFrenk2020}, the environment that surrounds a sub-critical halo with mass, $M_{200} \lesssim M_{\rm crit}$, impacts its inner gas content (see Eq. 5 in~\citetalias{BenitezLlambayFrenk2020}). Therefore, sub-critical haloes that form and evolve in denser environments have more gas at a fixed halo mass than similar haloes residing in underdense regions. The shaded region in the left panel of Fig.~\ref{Fig:Fig2_MgasMhaloAndSFRMhalo} shows the gas mass that results from applying the~\citetalias{BenitezLlambayFrenk2020} model but assuming a universe that is over/underdense by a factor of 3. We have verified in our simulation that the inner gas content of the haloes correlates with the mean density outside the haloes, indicating that the scatter in gas mass is due, to some extent, to the ambient pressure around the haloes (see Appendix~\ref{App:ambient_density}). The observed scatter around the median gas mass at fixed halo mass for sub-critical haloes is thus likely due to fluctuations of the ambient gas density relative to the mean.

The lowest panels of Fig.~\ref{Fig:Fig9_MassAccretionHistories} show that G3 and G4 undergo an initial burst of star formation after which both galaxies lose most of their baryons. As galaxy G3 inhabits a denser environment than G4, it can accrete more gas after the initial episode of star formation. At around $z\sim 3$, a further star formation event removes some gas from the galaxy. The gas mass of G3 finally stabilises at about $M_{\rm gas} \sim 5 \times 10^7 \ M_{\odot}$, a higher value than that expected for its halo mass. Note that G3 has maintained this configuration for roughly half of the Hubble time, implying that its current gas mass is not related to its past star formation, but to its environment, and the balance between the gas pressure and the halo gravity. Focusing on G4, we see that its initial star formation drives most of the baryons out of the system. Although some gas recovers secularly over a long timescale, the gas mass equilibrates at $M_{\rm gas} \sim 10^7$ at around $z\sim 1.5$ and remains constant until the present day. Therefore, as both galaxies have identical virial masses and have been quiescent for most of their lifetimes, their different present-day gas masses must stem from the different environments they inhabit today.

\subsection{Flyby galaxies}
\label{Sec:Flyby}

The clustering of gas-deficient galaxies around massive companions shown in Fig.~\ref{Fig:Fig3_OktantsHaloMassBin} together with the time evolution of G2 shown in the second row of Fig.~\ref{Fig:Fig4_4Examples} suggests that a significant fraction of gas-deficient galaxies must be flybys. 

As shown in Fig.~\ref{Fig:Fig9_MassAccretionHistories}, a distinctive feature of these flyby galaxies is that they lose both gas and dark matter as they approach a massive companion. Indeed, galaxy G2 has lost all its gas and $\sim 30$ per cent of its dark matter. 

Empirically, we find that flybys can be told apart from other galaxies in our simulation by requiring them to be present-day central galaxies that have lost dark matter in the past. Furthermore, we find that all these flybys were closer than $d<1.5 \times R_{200}$ of the eventual host, in good agreement with previous work~\citep[see, e.g.,][]{Behroozi2014}. Given these arguments, we define flyby galaxies in our simulation as present-day central systems that have lost more than 20 per cent of their dark matter and that were within $1.5 \times R_{200}$ of another galaxy at some point during their evolution. These criteria yield $328$ galaxies that, similarly to G2, have had a relatively strong interaction with a massive system in the past.

\begin{figure}
    \centering
    \includegraphics[width=\columnwidth]{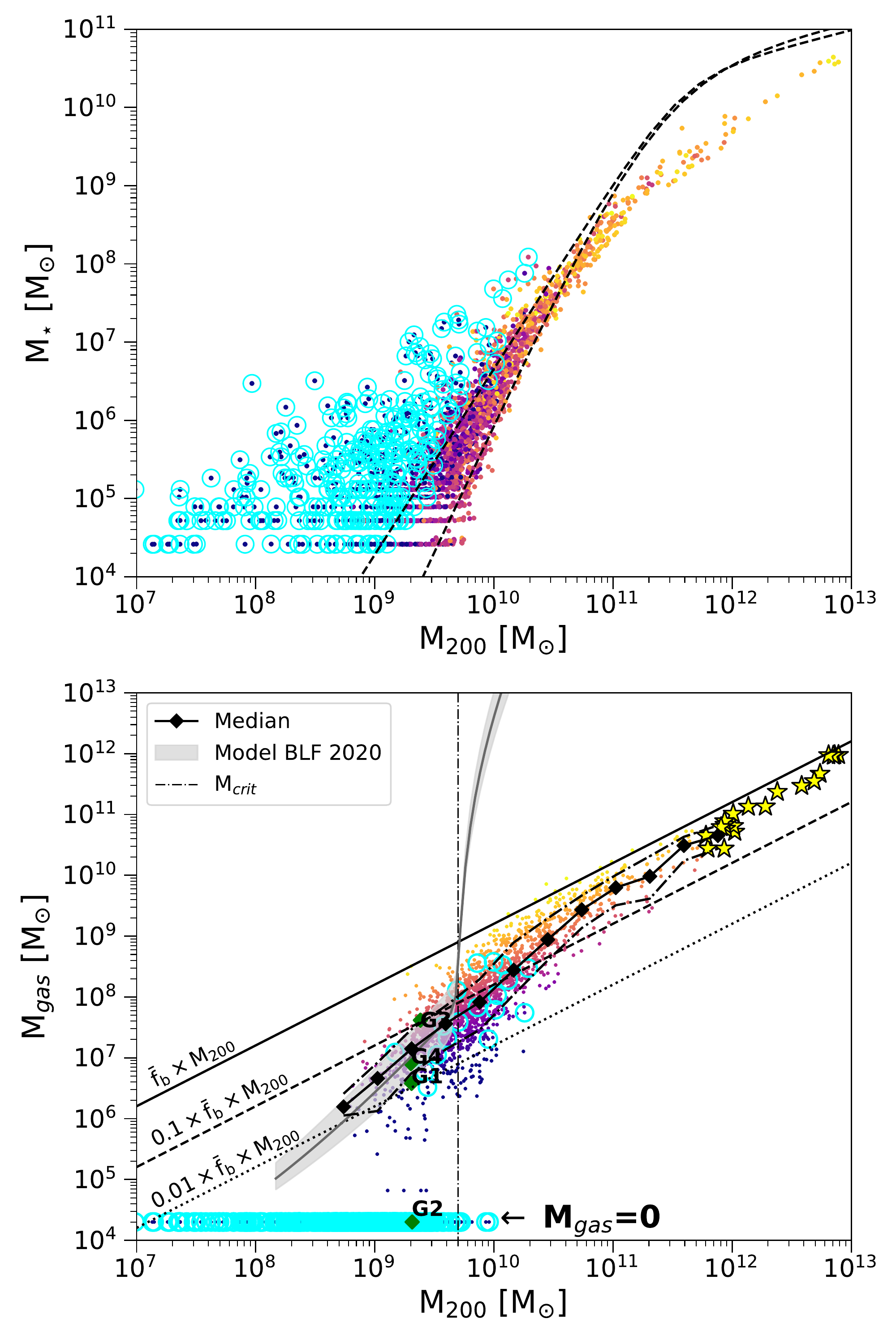}
    \caption{The top panel, analogous to Fig.~\ref{Fig:Stellar-vs-HaloMass}, shows the present-day stellar mass versus halo mass for our galaxy sample. The cyan circles indicate flyby galaxies, as defined in Sec.~\ref{Sec:Flyby}. Interestingly, flybys scatter off the stellar mass versus halo mass relation followed by non-flybys. The lower panel, analogous to Fig.~\ref{Fig:Fig2_MgasMhaloAndSFRMhalo}, shows the current gas mass as a function of halo mass. As in Fig.~\ref{Fig:Fig2_MgasMhaloAndSFRMhalo}, we show gas-free galaxies with an arbitrary gas mass value, $M_{\rm gas} = 2\times 10^4 \rm \ M_{\odot}$. The majority of flybys (cyan circles) are essentially devoid of gas.}
    \label{Fig:Fig8_MstarMhaloAndMgasMhalo_MarkBSwithCircles}
\end{figure}

\begin{figure*}
    \includegraphics[width=\textwidth]{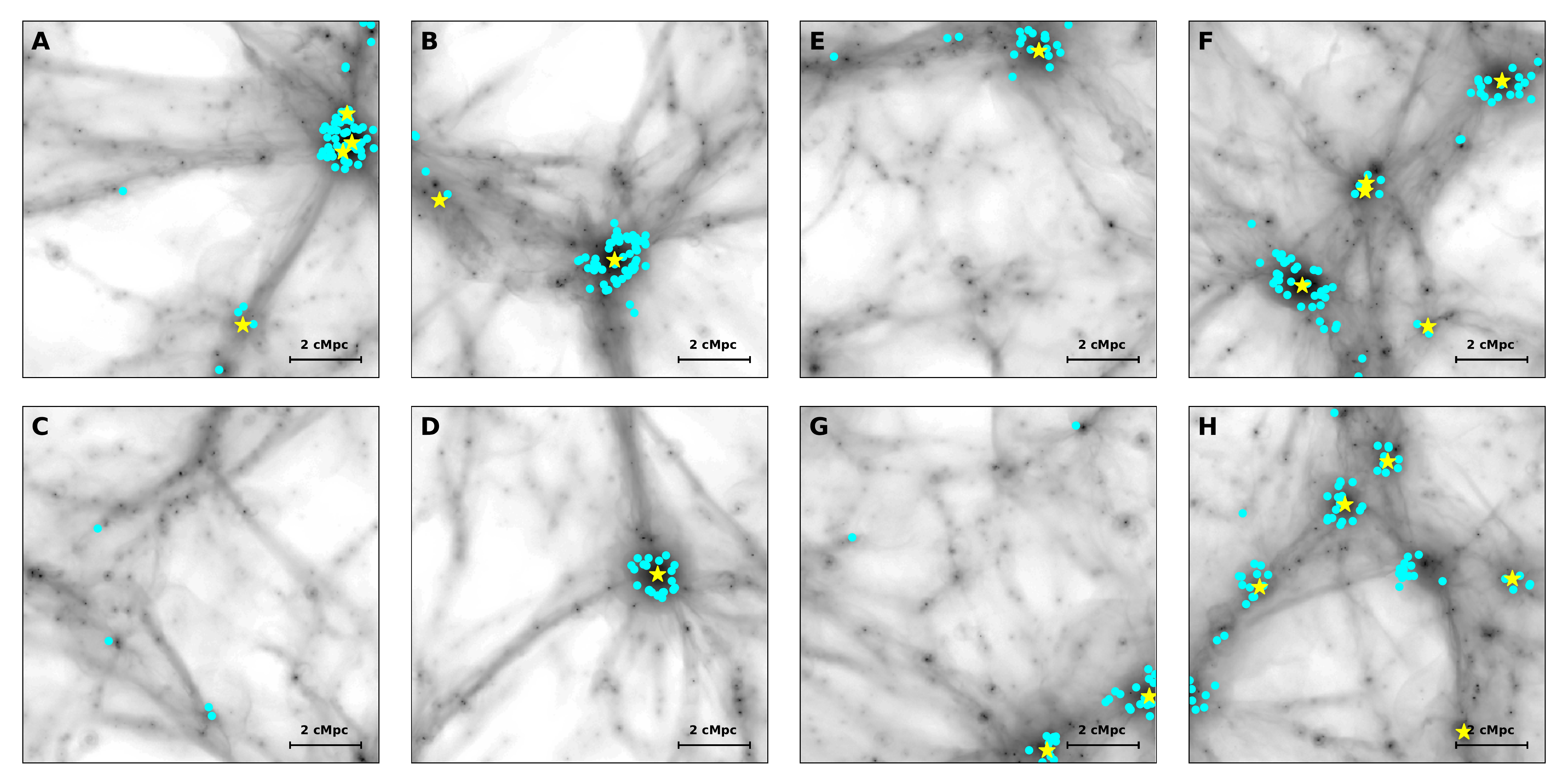} 
    \caption{Analogous to Fig.~\ref{Fig:Fig3_OktantsHaloMassBin}, but we now display the spatial distribution of flyby galaxies (cyan symbols), as defined in Sec.~\ref{Sec:Flyby}. Flyby galaxies exhibit a characteristic clustering pattern towards the more massive systems of the simulation (yellow stars).}

    \label{Fig:Fig6b_OktantsBS}
\end{figure*}

Fig.~\ref{Fig:Fig8_MstarMhaloAndMgasMhalo_MarkBSwithCircles} shows that our classification criteria have interesting consequences. The top panel of this figure shows the present-day gas mass as a function of halo mass for our galaxy sample. We show flyby galaxies with open cyan circles. Interestingly, at a fixed stellar mass flybys inhabit lower mass haloes than non-flyby counterparts. This dark matter deficit originates from the tidal interaction between the flybys and the eventual host. Secondly, as shown in the bottom panel of the same figure, most flybys have lost all the gas. Thirdly, flybys cluster today near massive systems, as shown in Fig.~\ref{Fig:Fig6b_OktantsBS}. This figure is analogue to Fig.~\ref{Fig:Fig3_OktantsHaloMassBin}, but we only display the location of the flyby galaxies that result from our selection criteria. Flybys depict well the location of the 20 most massive systems of our simulation, shown by yellow stars. Finally, a minority of flybys lie well beyond the virial radius of massive systems. For example, towards the top left corner of panel H in Fig.~\ref{Fig:Fig6b_OktantsBS} we find a flyby galaxy located beyond 2 Mpc from all massive hosts in the simulation, making it difficult to associate this flyby to a host from its present-day location. In a recent work,~\cite{Benavides2021} reported similar examples.

We thus conclude that a significant number of simulated isolated gas-deficient galaxies are flybys that have interacted with a massive companion in the past. We will quantify the contribution of these galaxies to the population of gas-deficient galaxies in Sec.~\ref{Sec:Discussion}.

\subsection{Cosmic-Web Stripped (COSWEB) galaxies}
\label{Sec:COSWEBs}

\begin{figure}
    \centering
    \includegraphics[width=\columnwidth]{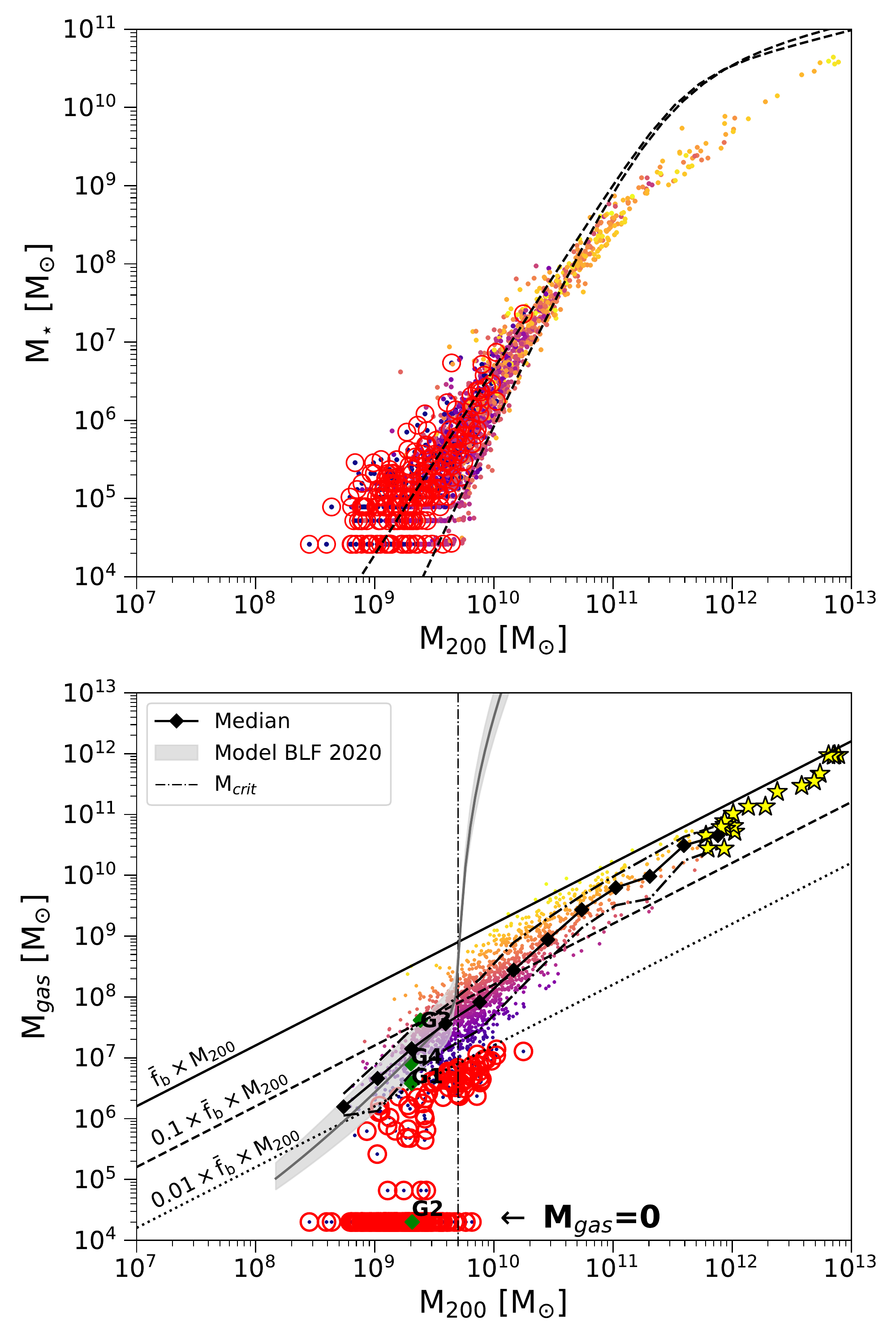}
    \caption{The top panel, analogous to Fig.~\ref{Fig:Stellar-vs-HaloMass}, but in which we removed flyby galaxies, shows the present-day stellar mass as a function of halo mass for our galaxy sample. The red circles indicate COSWEB galaxies, as defined in Sec.~\ref{Sec:COSWEBs}. The lower panel, analogous to Fig.~\ref{Fig:Fig2_MgasMhaloAndSFRMhalo}, shows the gas mass as a function of halo mass for all galaxies but flybys. As in the top panel, we show COSWEBs with red circles. Gas-free galaxies are indicated with a lower gas mass, 
    $M_{\rm gas} = 2\times 10^4 \rm \ M_{\odot}$.}
    \label{Fig:Fig8b_MarkCOSWEBs}
\end{figure}

Having established the reason behind the gas deficit for a significant fraction of the sample of gas-deficient galaxies, we now focus on the remaining gas-deficient galaxies that are not flybys. We show these systems with red open circles in Fig.~\ref{Fig:Fig8b_MarkCOSWEBs}. This figure, which is analogue to Fig.~\ref{Fig:Fig8_MstarMhaloAndMgasMhalo_MarkBSwithCircles} but in which we removed flybys, demonstrates that not all the gas-deficient galaxies are flybys. Indeed, after removing flybys, there is still a sizeable population of 231 gas-deficient systems that have not interacted with massive systems in the past. 
The gas deficit of most of these galaxies cannot be due to the effect of supernova feedback. Indeed, most of these systems inhabit sub-critical dark matter haloes, implying they have been quiescent for a long time (see right panel of Fig.~\ref{Fig:Fig2_MgasMhaloAndSFRMhalo}). These systems, albeit gas deficient today, are similar to other simulated systems in terms of their stellar mass (see the top panel). What is the reason behind the deficit of gas for these dwarfs?

\begin{figure*}
    \includegraphics[width=\textwidth]{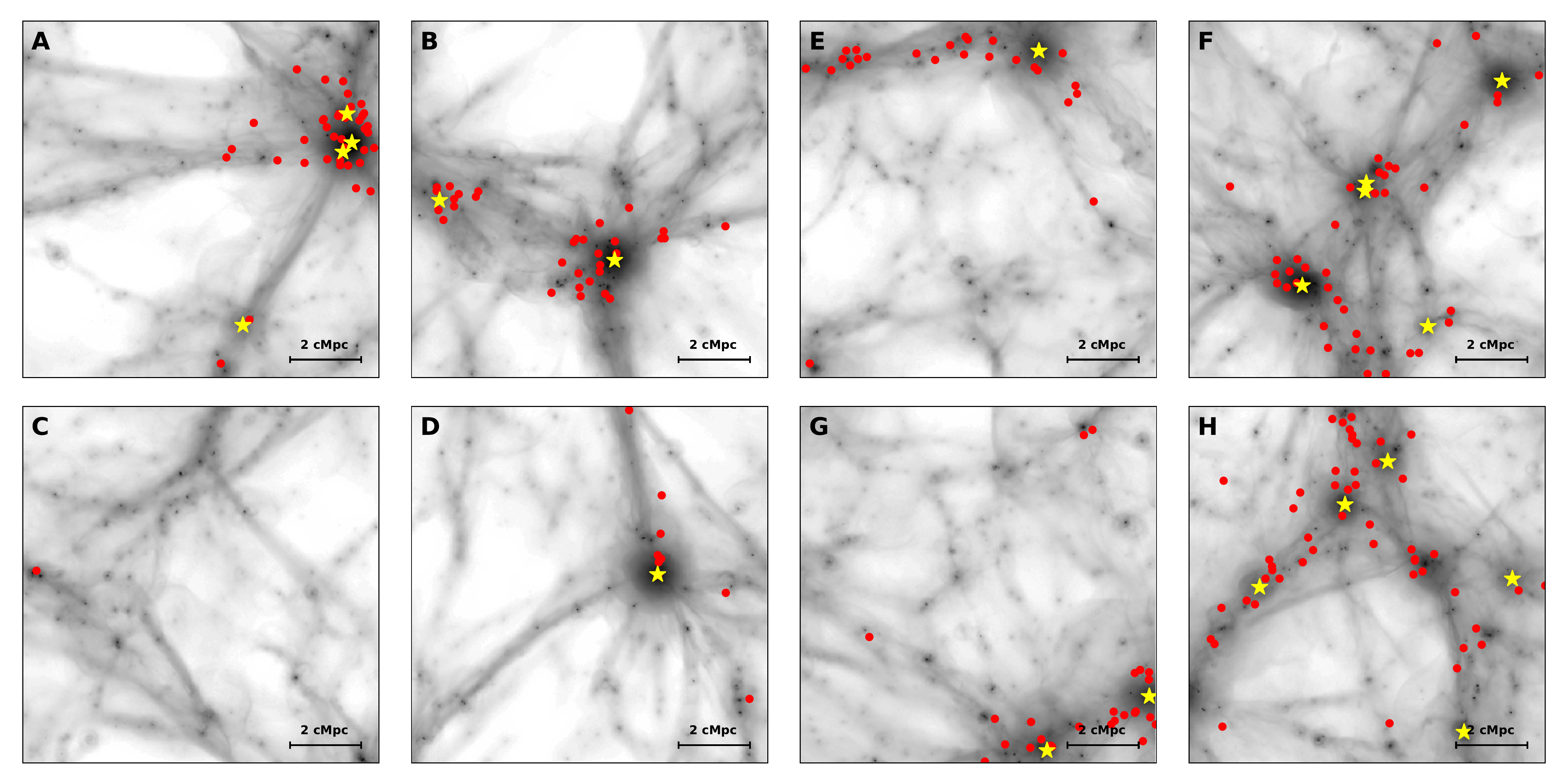} 
    \caption{Analogous to Fig.~\ref{Fig:Fig3_OktantsHaloMassBin}, but we now display the spatial distribution of COSWEB galaxies only (red symbols), as defined in Sec.~\ref{Sec:COSWEBs}. COSWEBs exhibit much less clustering towards massive systems (yellow stars) than flybys (see Fig.~\ref{Fig:Fig6b_OktantsBS}), and they depict the location of the filaments of the cosmic web well. We have verified that those COSWEB galaxies close to a massive system today are approaching the host for the first time. These galaxies have been stripped of their gas several virial radii away from the massive systems, so their gas removal is not directly related to ram-pressure exerted by the host's hot gaseous halo.}
    \label{Fig:Fig6a_OktantsBPNoBS}
\end{figure*}

Fig.~\ref{Fig:Fig6a_OktantsBPNoBS} shows the spatial distribution of the gas-deficient galaxies that are not flybys. Interestingly, these galaxies cluster less than flybys towards the massive systems of our simulation, and they predominantly depict the location of the filaments and sheets of the cosmic web (see, e.g., panels D, E, H).
Galaxies visually close to a massive host in our simulation are typically on their first infall, so they have not yet lost dark matter via tidal stripping. A visual inspection of the evolution of these systems through movies of their surrounding gas distribution made with the Py-SPHViewer code \citep{Benitez-Llambay2015}, together with the analysis of their individual mass assembly histories, revealed that they are all analogues of galaxy G1 (shown in the first row of Fig.~\ref{Fig:Fig4_4Examples}), i.e. these are systems that have lost their gas through hydrodynamic interactions with the cosmic web and denser gaseous structures which are not obviously associated with gaseous galaxy haloes. The only exception are ten galaxies that become gas-deficient due to SN feedback. We, therefore, conclude that our methodology captures the underlying physics responsible for removing gas from the system well. Because of this, we shall define cosmic-web-stripped (COSWEB) galaxies as gas-deficient galaxies that have lost their gas via ram pressure in the past and are far from other halos. This definition results in 221 COSWEB galaxies (i.e., all but 10 gas-deficient galaxies that are not flybys). 

\section{Discussion}
\label{Sec:Discussion}

In the previous sections, we demonstrated that the present-day population of field gas-deficient galaxies originates from either a close interaction with a massive host in the past (flybys) or ram-pressure stripping with the cosmic web (COSWEBs). The difference between flybys and COSWEBs becomes evident in their clustering properties. Flybys cluster towards the most massive galaxies present in our simulation. COSWEBs, on the other hand, are more dispersed throughout the simulation volume, preferentially depicting the location of the gaseous filaments and sheets of the cosmic web. What is the relative contribution of flybys and COSWEBs to the population of present-day gas-deficient galaxies, i.e., galaxies whose gas mass fraction is $M_{\rm gas}/M_{200} < 0.01\times \Omega_{b}/\Omega_{m}$?

\begin{figure}
    \centering
    \includegraphics[width=\columnwidth]{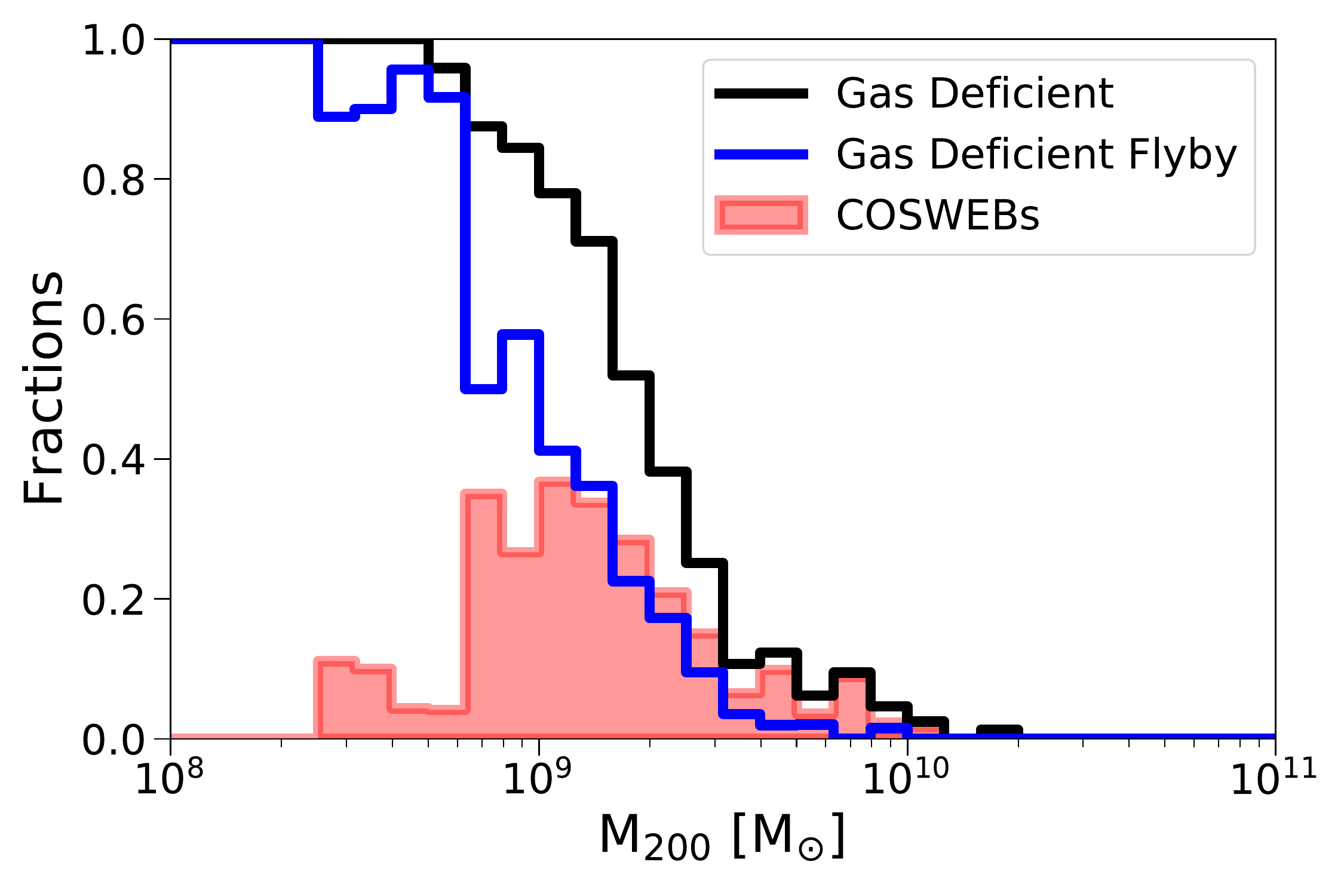}
    \caption{Fraction of gas-deficient (black line), flyby (blue line), and COSWEB (red area) galaxies, as a function of halo mass. We define gas-deficient galaxies as those whose virial gas mass fraction is under $1\%$ of the universal baryon fraction, $\bar f_{\rm b}=\Omega_{\rm b} / \Omega_{m}$. For consistency, we only considered those flyby galaxies that are also gas-deficient. We, therefore, excluded only 17 flybys that do not fulfil this requirement. Below a halo mass, $M_{200} \sim 5 \times 10^8 \rm \ M_{\odot}$, all galaxies become gas deficient (see text for a discussion on this). Below a halo mass, $M_{200} \sim 3\times 10^8$ M$_{\odot}$, all gas-deficient galaxies become flybys.}
    \label{Fig:Fig7_Fractions}
\end{figure}

In Fig.\ref{Fig:Fig7_Fractions} we show the fraction of gas-deficient galaxies (black line) as a function of halo mass. The fraction of gas-deficient galaxies reaches unity below the halo mass $M_{200}\lesssim 5 \times 10^{8} \ M_{\odot}$ because the gas mass of every dark matter halo eventually falls below $M_{\rm gas} < 0.01 \left (\Omega_{b}/\Omega_{m}\right ) M_{200}$, which is the maximum gas mass below which galaxies become gas-deficient in our definition. In addition, the limited resolution of our simulation makes the gas content of halos with mass $M_{200} \lesssim 10^9 \ M_{\odot}$ only marginally resolved. Indeed, the~\citetalias{BenitezLlambayFrenk2020} model indicates that the average gas mass of dark matter halos of mass $M_{200} \sim 10^{9} \ M_{\odot}$ is $M_{\rm gas} \sim 10^6 \ M_{\odot}$, or $\sim 20$ gas particles in our simulation, so we caution not to overinterpret results in this regime. However, for masses $M_{200} > 10^{9} \ M_{\odot}$, we expect the fraction of gas-deficient galaxies to be robust.

Next, we quantify the relative contribution of flybys and COSWEBs to the population of present-day gas-deficient galaxies. We show this in Fig.~\ref{Fig:Fig7_Fractions}, where the fraction of flybys and COSWEBs, relative to the population of gas-deficient galaxies, are indicated by the blue and red histograms, respectively. COSWEBs dominate over flybys at higher masses. However, less than 10 per cent of the field galaxies (but essentially all gas-deficient galaxies) are COSWEBs today for halo masses $M_{200} \gtrsim 5 \times 10^{9} \ M_{\odot}$, as only a negligible fraction of galaxies are flybys within the same mass range. The lack of massive flybys is not surprising, as these galaxies, unlike COSWEBs, have all lost dark matter, thus biasing the population of flybys towards lower masses. The contribution of COSWEBs peaks at around $M_{200} \sim 10^{9} M_{\odot}$, making up roughly 35 per cent of the field galaxies (and half of the gas-deficient galaxies) at that mass before vanishing below a halo mass $M_{200} \lesssim 3 \times 10^8 \ M_{\odot}$,  which is the limit where galaxies stop forming in our simulation~\citep{BenitezLlambayFrenk2020, BenitezLlambay2021}. In contrast, the contribution of flybys becomes more relevant with decreasing halo mass, which is not surprising. Indeed, galaxies can form in our simulation only in haloes that exceed the atomic cooling limit, which imposes a minimum halo mass to host a galaxy today of $M_{200} \gtrsim 3 \times 10^{8} \ M_{\odot}$. The only way a galaxy can inhabit a halo less massive than this critical mass in our simulation is by decreasing the halo mass, as is the case for flybys.

The details of our analysis may be influenced, to some extent, by the specifics of our definitions or the modelling included in our simulation. Indeed, some of our COSWEB galaxies could be classified as flybys if we made our classification criteria less stringent. However, this would only reduce the number of COSWEBs galaxies in our simulation. Thus, we can safely and robustly conclude that cosmic web stripping affects the gas content of less than 10 per cent of the simulated galaxies that would otherwise be able to foster star formation. These are galaxies with virial mass $M_{200} \gtrsim 5 \times 10^9 \ M_{\odot}$, above which gas, if available, can collapse and form stars. Also, although cosmic web stripping is a rare process for massive dwarfs, it dramatically affects the gas fraction and the ability to form stars in most affected dwarfs. We have verified that more than 70 per cent of massive COSWEBs that inhabit haloes that could sustain star formation today, i.e. those with  $M_{200} \gtrsim 5 \times 10^9 \ M_{\odot}$, are quiescent at the present day due to the lack of gas. These are the systems shown with zero SFR to the right of the critical mass (vertical line) in the right panel of  Fig.~\ref{Fig:Fig2_MgasMhaloAndSFRMhalo}. It is interesting to see that essentially none of the massive COSWEB galaxies lose all the gas after interacting with the cosmic web. In all cases, a small amount of gas remains in the centre. This remnant originates after cosmic web stripping removes the outer low-density gas, causing the gradual expansion of the inner dense star-forming gas. Thus, the lack of a cold molecular phase in our simulation does not prevent the gas from reaching high enough densities and resisting the ram pressure. However, assessing the importance of the cold phase to the present-day properties of the COSWEB dwarfs requires dedicated simulations that include the explicit treatment of molecular cooling.

The scarcity of massive COSWEBs contrasts sharply with the 80 per cent fraction found by~\citetalias{Benitez-Llambay2013} within a similar halo mass range in a simulation of the formation of the Local Group. The low fraction of massive COSWEBs in our simulation thus indicates that the effect of cosmic web stripping is small over cosmological scales, affecting roughly 1 in 10 massive dwarfs in the field. We note, however, that the~\citetalias{Benitez-Llambay2013} results may indicate that cosmic web stripping is a more frequent process nearby biased regions, such as the Local Group, whereas our results asses the average importance over cosmological scales. Fig.~\ref{Fig:Fig6a_OktantsBPNoBS}, together with Fig.~\ref{Fig:Fig6b_OktantsBS} and Fig.~\ref{Fig:Fig3_OktantsHaloMassBin}, reveal that the sheer number of COSWEBs and flybys really depends on the environment. However, we do not find any region in our simulation containing a fraction of COSWEB as high as that found by~\citetalias{Benitez-Llambay2013}. Therefore, we believe that the discrepancies largely arise from the biased region studied by these authors, which is constrained and departs from a random $\Lambda$CDM realisation, together with the reduced number of dwarfs they consider.

Also, it is interesting to see that cosmic web stripping can promote star formation in some systems, at least for some time. We do not quantify this, but our simulation contains example dwarfs inhabiting haloes with mass under the critical mass whose gas is compressed by the cosmic web, thus triggering star formation. Galaxy G1 is an example (see Fig.~\ref{Fig:Fig9_MassAccretionHistories}). Using simulations of the formation of the Local Group,~\citet{Pereira-Wilson2022} have recently pointed out similar examples. These results agree qualitatively with~\citet{Wright2019}, who find that interactions with the cosmic web can promote star formation in dwarf galaxies.

For less massive dwarfs, cosmic web stripping becomes a more frequent process. However, we expect it to have only a minor impact on the galaxy, as these galaxies have already lost most of their gas due to the effect of the UVB radiation field. The remaining small amount of gas is in hydrostatic equilibrium and unable to condense to the centre to form stars. Although the loss of gas in these low-mass galaxies has little importance for their evolution, it may have observational implications for their detection in the next-generation HI surveys. \citet{BenitezLlambay2017} have shown that these small haloes can contain HI masses $M_{\rm HI} \gtrsim 10^3 \ M_{\odot}$, and reach column densities $N_{\rm HI} \gtrsim 10^{18} \rm \ cm^{-2}$. This value is below current threshold limits for detectability but may well be reached in the future with upcoming instruments.

\section{Summary and Conclusions}
\label{Sec:Conclusions}

We examined the gas content of simulated field galaxies identified in a high-resolution cosmological hydrodynamical simulation. We showed that the simulated galaxies are naturally split into quiescent and star-forming (those that have/have not formed stars during the last $2 \ \rm Gyrs$, respectively) based on how the mass of their haloes compares to a well-understood critical mass. In agreement with previous work, we find that galaxies residing in haloes more massive than the present-day value of the ~\citetalias{BenitezLlambayFrenk2020} critical mass are largely star-forming, whereas those galaxies inhabiting lower mass haloes are quiescent.

Of the quiescent galaxies, we find that those that contain gas today follow the~\citetalias{BenitezLlambayFrenk2020} remarkably well. This indicates that, on average, the gas mass of luminous galaxies that inhabit sub-critical dark matter haloes depends on the balance between the gas pressure and the halo gravity and not on feedback from evolving stars. This fact is not minor, as this allowed us to use the~\citetalias{BenitezLlambayFrenk2020} model to target galaxies with a reduced gas content while ruling out supernova feedback as the main culprit for their gas deficit. These are the only galaxies whose current gas content could have been severally affected by an external process. 

Our conclusions may be summarised as follows:

\begin{itemize}
    \item{Galaxies inhabiting sub-critical dark matter haloes, namely haloes with mass, $M_{200} \lesssim M_{\rm crit} \sim 5 \times 10^{9} \ M_{\odot}$, are quiescent today, and they have been so for a long time. This is because the gas in these haloes is in hydrostatic equilibrium and unable to undergo gravitational collapse and form stars in the centre. The gas mass of the galaxies that inhabit these low-mass haloes is thus well understood and can be calculated in detail by the simple~\citetalias{BenitezLlambayFrenk2020} model (see Fig.~\ref{Fig:Fig2_MgasMhaloAndSFRMhalo}).}
    \item{We find that the ambient density around our galaxy sample is of the same order as the fluctuations in the galaxys' gas mass (at a fixed halo mass). This indicates that the scatter in gas mass for sub-critical haloes at fixed halo mass is due to the environment and unrelated to past events of star formation (see Appendix A, Fig.~\ref{Fig:Fig4_4Examples} and Fig.~\ref{Fig:Fig9_MassAccretionHistories}).}
    \item{A non-negligible number of galaxies inhabiting sub-critical haloes today are gas-deficient, meaning their gas mass falls well below the value (and the scatter) expected for their halo mass. The exacerbated gas deficit in these galaxies cannot stem from the effect of supernova feedback or the external UVB. The origin of these gas-deficient galaxies relates to their present-day environment and their past evolution (see Fig.~\ref{Fig:Fig3_OktantsHaloMassBin} and Fig.~\ref{Fig:Fig4_4Examples}).}
    \item{Most simulated gas-deficient galaxies originate from past interactions with the most massive galaxies that form in our volume. Although our galaxy sample considers central galaxies today, we find that many field galaxies are, in fact, ``flybys'' that lost their gas (and also dark matter) in past interactions with massive hosts. Most simulated flybys do not contain gas today, and they cluster towards the most massive galaxies of our volume (see, e.g., Fig.~\ref{Fig:Fig4_4Examples} and Fig.~\ref{Fig:Fig6b_OktantsBS}).}
    \item{A substantial number of gas-deficient galaxies are not flybys. Close inspection of their evolution and clustering properties reveals that these systems have lost their gas through hydrodynamic interactions with the gaseous filaments and sheets of the cosmic web. We refer to these as cosmic-web stripped (COSWEBs) galaxies (see Fig.~\ref{Fig:Fig8b_MarkCOSWEBs}).}
    \item{Flyby galaxies make up 100 per cent of the gas-deficient galaxies below the present-day halo mass of $M_{200} \lesssim 3 \times 10^{8} \ M_{\odot}$, as galaxy formation only proceeds in haloes more massive than the atomic cooling limit in our simulation. Galaxies inhabiting haloes less massive than this limit must have formed in haloes that underwent heavy dark matter stripping in the past, as flybys do.}
    \item{COSWEBs are more frequent than flybys at high halo masses ($M_{200}\gtrsim 5 \times 10^{9} \ M_{\odot}$). This mass dependence originates from the loss of dark matter by flybys. We find that cosmic web stripping affects less than 10 per cent of the simulated galaxies that could otherwise sustain star formation today. These are galaxies that inhabit dark matter haloes more massive than the critical mass, $M_{200} \gtrsim M_{\rm crit}\sim 5 \times 10^{9} \ M_{\odot}$. Cosmic web stripping thus affects, on average, only a small fraction of the cosmological population of star-forming dwarfs.}
    \item{More than 70 per cent of COSWEB galaxies inhabiting haloes with mass $M_{200}>M_{\rm crit}$ are quiescent today. This indicates that cosmic web stripping, albeit of low frequency, is an efficient process at shutting off star formation in dwarfs that would otherwise be able to form stars today.}
    \item{The fraction of COSWEB galaxies increases for sub-critical haloes, peaking at $\sim 35$ per cent at $M_{200} \sim 10^{9} \ M_{\odot}$. The small gas content of galaxies inhabiting these sub-critical haloes cannot collapse to make stars in their centre, so cosmic web stripping has a negligible impact on the properties of these galaxies. Although the loss of gas in these low-mass galaxies has little importance for their evolution and present-day properties, it may have observational implications for their detection in upcoming HI surveys. Simple analytic and numerical considerations~\cite[see e.g., ][]{BenitezLlambay2017}, indicate that these galaxies may contain over $10^3-10^{4} \ M_{\odot}$ of neutral hydrogen, and reach column densities, $N_{\rm HI} \gtrsim 10^{18} \rm \ cm^{-2}$. The removal of gas in these low-mass systems may thus add up to the expected diversity in the HI mass of dwarf galaxies~\citep{Rey2022}, and preclude the future detection of a large number of faint dwarfs in future blind HI surveys. }
\end{itemize}

 Our analysis demonstrates, for the first time, that cosmic-web stripping affects only a low fraction of dwarfs massive enough to sustain star formation at the present day. The classification of gas-deficient galaxies into flybys or COSWEBs becomes ambiguous for galaxies that have not undergone strong gravitational interactions with a host in the past, or for those that may be approaching a massive companion for the first time today. Our estimates of the importance of cosmic-web stripping thus constitute an upper limit of the significance of this process over cosmological scales. 
 Although the exact numbers we derive in our analysis may be sensitive to the particularities of our definitions, we note that the existence of a significant number of gas-deficient galaxies is a robust result of our analysis. Our study reveals the important role of the environment in shaping the gas content of field dwarf galaxies over cosmological scales.

\section*{Acknowledgements}
We thank the referee, Dr Martin Rey, for constructive feedback. This work is supported by the European Research Council (ERC) under the European Union's Horizon 2020 research and innovation program under the grant agreements 101026328 and 757535, and UNIMIB's Fondo di Ateneo Quota Competitiva (project 2020-ATESP-0133). This work used the DiRAC Data Centric system at Durham University, operated by the Institute for Computational Cosmology on behalf of the STFC DiRAC HPC Facility (https://www.dirac.ac.uk). This equipment was funded by BIS National E-infrastructure capital grants ST/P002293/1, ST/R002371/1, and ST/S002502/1, Durham University, and
STFC operations grant ST/R000832/1. DiRAC is part of the National e-Infrastructure. The simulation used in this work was performed using DiRAC’s director discretionary time awarded to ABL. ABL is grateful to Prof. Mark Wilkinson for awarding this time.

\section*{Data Availability}
The simulation data underlying this article will be shared on reasonable request to the corresponding author.



\bibliographystyle{mnras}
\bibliography{CosmicWebStripping} 



\appendix
\section{Environment}
\label{App:ambient_density}

\begin{figure}
    \centering
    \includegraphics[width=\columnwidth]{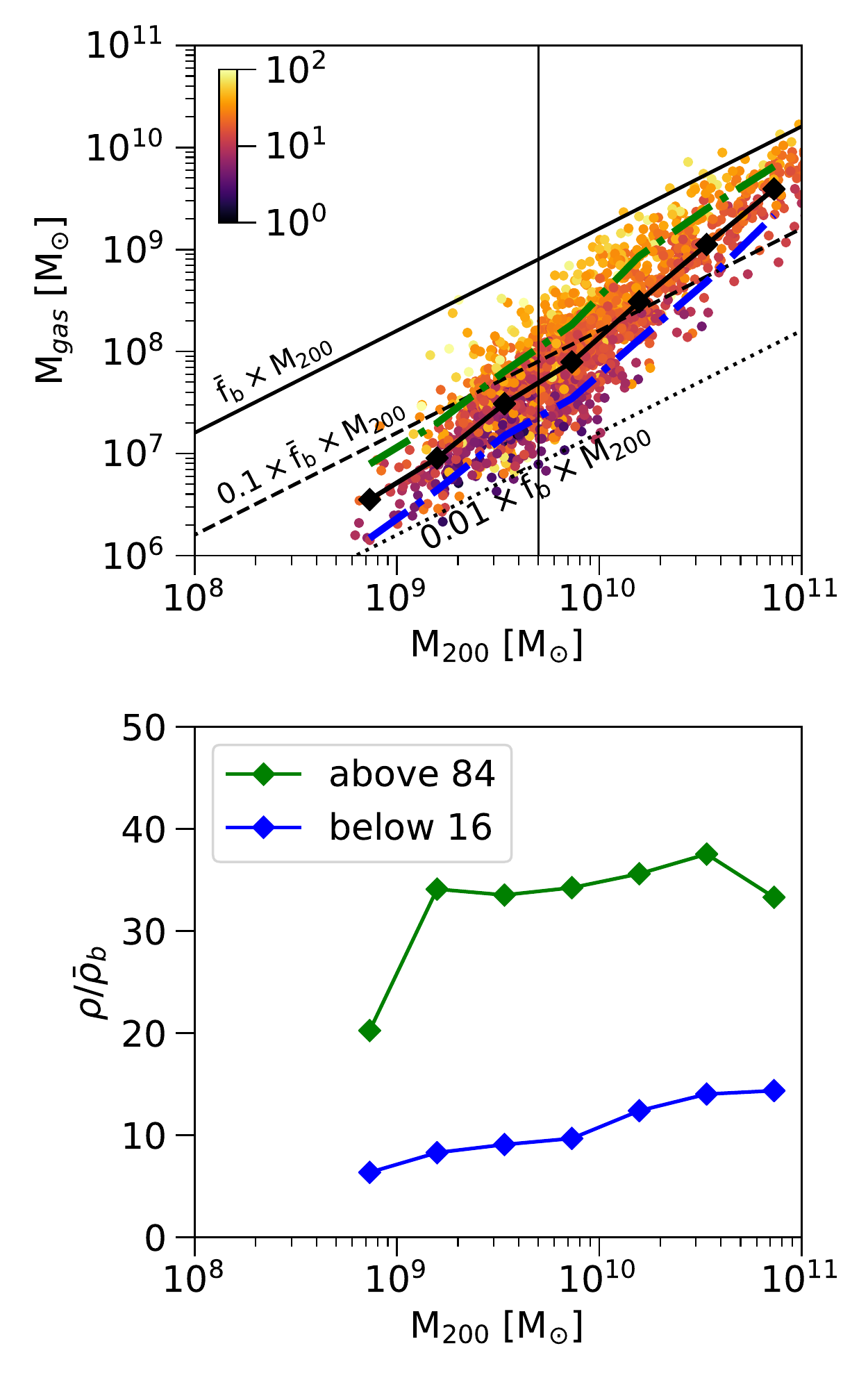}
    \caption{The top panel shows the gas mass as a function of halo mass for our galaxy sample, excluding gas-deficient galaxies. Galaxies are coloured according to their ``ambient'' density, i.e., the mean gas density within a shell between 2 to 3 times the virial radius of the systems, in units of the mean density of the Universe. The coloured lines show the median and the 16-84th percentiles of the distribution. The oblique lines show the expected gas mass for haloes that have retained the universal baryon fraction (solid line), 10\% of this value (dashed line), and 1\% (dotted line). The vertical line shows the~\citetalias{BenitezLlambayFrenk2020} critical mass. The bottom panel shows the median ``ambient'' density relative to the mean baryon density of the Universe, as a function of halo mass, for galaxies whose gas mass is above the 84th percentile (green line), and for galaxies whose gas mass is under the 16th percentile (blue line). The difference in ``ambient'' density between gas-rich and gas-poor galaxies indicates that the gas mass of these halos is affected, to some extent, by the amount of gas located beyond the virial radius of the systems.}
    \label{Fig:FigAppendix}
\end{figure}

To investigate the reason for the scatter in gas mass for galaxies that are neither flybys nor COSWEBs, we now consider the ``ambient'' gas density. We define the ``ambient'' gas density as the mean density of a shell centred at each galaxy, and located between $2\times R_{200}$ to $3\times R_{200}$. The upper panel of Fig.~\ref{Fig:FigAppendix} shows the gas mass, as a function of halo mass, for our galaxy sample, colouring galaxies according to their ``ambient'' density relative to the mean baryon density of the Universe. Interestingly, at fixed halo mass, galaxies with higher gas mass display higher values of ``ambient'' density, whereas the opposite is true for galaxies with low gas mass. 

In the lower panel of Fig.~\ref{Fig:FigAppendix}, we quantify the difference in the ``ambient'' gas density between gas-rich and gas-poor galaxies as a function of halo mass. To this end, we display the median ``ambient'' density relative to the mean baryon density of the Universe for the galaxies with gas mass above the $84$th percentile (green line in the top panel), and for those with gas masses below the $16$th percentile (blue line in the top panel). The median ``ambient'' density for galaxies with a high gas mass almost triples that of low gas mass galaxies. This factor, albeit smaller than the scatter in gas mass, indicates that the gas mass within the haloes is determined, to some extent, by the environment that surrounds the haloes, as expected from the simple arguments of~\cite{BenitezLlambayFrenk2020}.

\bsp	
\label{lastpage}
\end{document}